\documentclass[usegraphicx,usenatbib]{mn2e}
\usepackage{times}
\def\mpc{h^{-1} {\rm{Mpc}}}
\def\kms {\rm{km~s^{-1}}}
\def\apj {ApJ}
\def\apjl {ApJL}
\def\apjs {ApJS}
\def\aj {AJ}
\def\aap {A\&A}
\def\mnras {MNRAS}
\begin{document}
\title[Galaxies in clusters]
{On the relationship between environment and galaxy properties in clusters
of galaxies}
\author[H.J. Mart\'{\i}nez, V. Coenda \& H. Muriel]
{H\'ector J. Mart\'\i nez,\thanks{E-mail: julian@oac.uncor.edu} 
Valeria Coenda \& Hern\'an Muriel\\
Instituto de Astronom\'{\i}a Te\'orica y Experimental,
IATE, CONICET$-$Observatorio Astron\'omico, Universidad Nacional de 
C\'ordoba,\\ Laprida 854, X5000BGR, C\'ordoba, Argentina}
\date{\today}
\pagerange{\pageref{firstpage}--\pageref{lastpage}} 
\maketitle
\label{firstpage}
\begin{abstract}
We study the correlation between different properties of bright ($L>L^{\ast}$) 
galaxies in clusters and the environment in the Sloan Digital Sky Survey 
(SDSS). Samples of clusters of galaxies used in this paper are those selected 
by Coenda \& Muriel that are drawn from the Popesso et al. and Koester 
et al. samples. Galaxies in these clusters have been identified in the Main 
Galaxy Sample of the Fifth Data Release (DR5) of SDSS. We analyse which galaxy 
properties correlate best with either, cluster mass or cluster-centric 
distance using the technique by Blanton et al. We find that galaxy properties 
do not clearly depend on cluster mass for clusters more massive than 
$M\sim10^{14}M_{\odot}$. On the other hand, galaxy properties correlate 
with cluster-centric distance. The property most affected by the 
cluster-centric distance is $g-r$ colour, closely followed by the $u-r$ 
colour. These results are irrespective of the cluster selection criteria. 
The two samples of clusters were identified based on the X-ray emission and 
the galaxy colours, respectively. Moreover, the parameter that best predicts 
environment (i.e. cluster-centric distance) is the same found by Mart\'\i nez 
\& Muriel for groups of galaxies and Blanton at al. for the local density 
of field galaxies.
\end{abstract}
\section{Introduction} 
Galaxy properties depend on environment, the high fraction of early type 
galaxies in rich cluster being one of the best examples. The fact that this 
fraction also evolves with time, i.e., an increasing of the S/S0 rate with 
redshift \citep{dressler97,fasano00}, strongly suggests that galaxy 
morphologies are being altered by the physical processes that act in the 
cluster environment. Depending on the latter, i.e., low or high local density, 
different physical mechanisms will affect galaxy properties in different ways. 
The fact that most galaxy properties are correlated (morphology, luminosity, 
colours, etc), makes it difficult to know which properties are affected most 
by the environment. \citet{blanton05} developed a test to evaluate which 
property, or pair of properties, are most predictive of the local density 
using galaxies in the Sloan SDSS \citep{sdss}. \citet{hm2} extended the 
analysis to galaxies in groups correlating galaxy properties with both, the 
mass of groups and the position in the system. \citet{blanton05} and 
\citet{hm2} found that galaxy colour is the property most predictive of the 
environment. This is particularly surprising taking into account the 
significant differences between field and group environments.

In high mass systems, the hot intracluster medium (ICM) becomes important. 
Mechanisms like ram pressure stripping \citep{gg72} and 
starvation/strangulation can affect both the gaseous content of galaxies 
\citep{abadi99} and the star formation history \citep{fujita99}. From the 
dynamical point of view, high speed encounters between galaxies are more 
frequent, producing morphological transformations \citep{moore98,moore99}.
Galaxies can also suffer the stripping  of gas and stars due to the 
interaction with the cluster potential (e.g. \citealt{moore99}).  It is not 
clear which of these, or some other proposed mechanisms, are dominant. The 
fact that different processes affect different galaxy components or 
properties, means that the correlation between these properties can also 
vary with the environment. Therefore, the most predictive galaxy properties 
could also depend on the type of system considered. \citet{yo02},
\citet{hm2}, \citet{weinmann06} and \citet{zmm06} found a clear dependence 
between the galaxy properties and the mass of the host group. For high mass 
clusters, the correlation is not clear. Several studies did not find 
dependence between the star formation rate or the fraction of early types with
masses or velocity dispersions of clusters (see for instance \citealt{goto03}).
Nevertheless, \citet{goto05} and \citet{margo01} found indications of a 
correlation between blue galaxy fraction with  the cluster richness. More 
recently, \citet{hansen07} have found that the fraction of red galaxies 
increases with the cluster mass, although only weakly for cluster more massive
than $\sim 10^{14} M_{\odot}$. \citet{popesso07} found that a luminous X-ray 
intracluster medium can affect the colour of galaxies. On the other hand, the 
local density-morphology \citep{huhu31,oemler74,dressler80} or the 
cluster-centric distance-morphology relation \citep{whitmore91} has been 
confirmed by several authors in different environments (see 
\citealt{dominguez01} for a discussion between these two approaches). 

In this paper we systematically explore the ability of different galaxy 
properties to predict the total mass of the cluster and the normalised 
cluster-centric distance. We have considered two samples of high mass clusters 
based on different selection criteria. 

This paper is organised as follows: in section 2 we describe the sample of 
galaxies in clusters; the analyses of the dependence of galaxy properties on 
mass and on the cluster-centric distance are carried out in sections 3 and 4
respectively. We summarise our results and discuss them in section 5.

Galaxy magnitudes used throughout this paper have been corrected for Galactic 
extinction using the maps by \citet{sch98}, absolute magnitudes have been 
computed assuming a flat cosmological model with parameters $\Omega_0=0.3$, 
$\Omega_{\Lambda}=0.7$ and $H_0=100~h~{\rm km~s^{-1}~Mpc^{-1}}$ 
and $K-$corrected using the method of \citet{blanton03}~({\small KCORRECT} 
version 4.1). All magnitudes are in the AB system.
\section{The sample of galaxies in clusters}
\subsection{The cluster sample}
Clusters of galaxies used in this paper has been taken from the cluster 
catalogue constructed by \citet{coenda08}. This catalogue was drawn from two 
cluster catalogues based on SDSS: ROSAT-SDSS Galaxy Cluster Survey of 
Popesso et al. (2004, hereafter P04), which is a X-ray selected cluster sample
and the MaxBGC Catalogue of Koester et al. (2007b, hereafter K07), which is an
optically selected cluster sample. Briefly, the ROSAT-SDSS catalogue provides 
X-ray properties of the clusters derived from the ROSAT data, and optical 
parameters computed from SDSS data. P04 includes clusters with masses from 
$10^{12.5}M_{\odot}$ to $10^{15}M_{\odot}$ in the redshift range 
$0.002\leq z \leq 0.45$. On the other hand, the optical MaxBGC catalogue relies
on the observation that the galaxy population of rich clusters is dominated by
the bright red galaxies tightly clustered in colour (the E/S0 ridgeline). The 
K07 catalogue comprises galaxy clusters with velocity dispersions 
$\sigma \ge 400 \kms$ and redshifts $0.1\le z\le 0.3$. 

The subsamples from P04 and K07 selected by \citet{coenda08}, labelled
as C-P04 and C-K07, comprise galaxy clusters in the redshift range 
$0.05<z<0.14$. For the K07 they also applied a restriction in the richness 
selecting clusters with $N_{\rm gal}\geq 20$ in order to have cluster masses 
comparable to those in the P04 sample. To select clusters members and estimate
the physical properties of clusters they used the Main Galaxy Sample (MGS; 
\citealt{mgs}) of the Fifth Data Release (DR5) of SDSS \citep{dr5} that is 
complete down to a \citet{petro76} magnitude $r=17.77$. To identify cluster 
members, \citet{coenda08} use the friends-of-friends (\textit{fof}) algorithm 
developed by Huchra \& Geller (1982) with percolation linking length values 
according to D\'iaz et al. (2004). As a result, they get for each field a list
of substructures with at least 10 members identified by \textit{fof}. The 
second step consists in a eyeball examination of the structures detected by 
\textit{fof}, a comparison between them and the listed cluster position and 
redshift to determine which coordinates and redshift fit best the observed 
galaxy over-density, i.e., the cluster centre. According to \citet{coenda08}, 
for $\sim 40\%$ of the clusters the angular position of the centre given by 
\textit{fof} is better than the original value, whereas for $\sim 17\%$ of the 
clusters the redshift according to \textit{fof} is a better match to the 
observed distribution than the listed one. From the redshift distribution of 
galaxies within $|cz-cz_{\rm cluster}|\le 3000\kms$ the authors determine the 
line-of-sight extension of the cluster, i.e., a maximum and a minimum redshift
for the cluster. They then consider as cluster members all galaxies in the 
field that lie within that redshift range.

Through visual inspection \citet{coenda08} classified clusters based on their 
substructure. For the purposes of this work, we will only consider the 
subsamples C-P04-I and C-K07-I that comprise regular clusters (type I in 
\citealt{coenda08}) and exclude systems that have two or more close 
substructures of similar size in the plane of the sky and/or in the redshift 
distribution. 

Once the members of each cluster are selected \citet{coenda08} compute some 
cluster physical properties we are interested in. Namely, they compute the 
line-of-sight velocity dispersion $\sigma$, virial radius and mass and the 
radius which encloses a mean over-density 200 times the mean density of the 
universe, $r_{200}$.

The mean values of these parameters are shown in table \ref{tb:mean} where it 
can be seen that the subsample drawn from P04 includes on average clusters 
slightly more massive and larger than the subsample taken from K07. The C-P04-I
and C-K07-I galaxy cluster samples comprise 49 and 209 clusters respectively.
\begin{table}
\center
\caption{Mean values of the cluster physical properties of our cluster 
samples.}
\begin{tabular}{ccccc}
\hline
           & $\sigma$ & $R_{200}$ & $M_{vir}$        & $R_{vir}$ \\
           & [$\kms$] & [$\mpc$]  & [$h^{-1}M_{\odot}$] &[$\mpc$]   \\
\hline
C-P04-I Sample & $715$    & $1.77$    & $7\times10^{14}$      & $1.75$    \\
\hline
C-K07-I Sample & $675$    & $1.67$    & $6\times10^{14}$       & $1.59$    \\
\hline
\end{tabular}
\label{tb:mean}
\end{table}
\begin{table}
\caption{Parameters' cut-offs that define our samples of galaxies in clusters} 
\begin{tabular}{lrr}
\hline
Property  & Minimum Value & Maximum Value \\
\hline
$M_{^{0.1}r}-5\log(h)$  &  $-22.5$  &  $-20.4$ \\ 
$^{0.1}(u-r)$           &  $1.7$    &  $3.2$   \\
$^{0.1}(g-r)$           &  $0.55$   &  $1.1$   \\
$\mu_{50}$                 &  $19.25$  &  $22$    \\
$C=r_{90}/r_{50}$       &  $1.9$    &  $3.5$   \\ 
$\log(n)$               &  $-0.5$   &  $1.0$   \\
eclass                  &  $-0.25$  &  $0.1$   \\
\hline
\label{cutoffs}
\end{tabular}
\end{table}
\subsection{Galaxy parameters}
As in \citet{hm2} and to avoid systematic effects, we have constructed volume 
limited samples of galaxies instead of using flux limited samples with a galaxy
weighting scheme to account for Malmquist bias. This is crucial for a fair 
comparison of galaxies in clusters at different redshifts. Thus we basically 
deal with galaxies brighter than $M_{^{0.1}r}-5\log(h)=-20.4\simeq M^{\ast}$ 
and $z<0.14$. With this restriction our samples of galaxies comprise: 786 
galaxies from the C-P04-I and 3041 from the C-K07-I samples.

Among the available data for each object in the MGS, we have used in our 
analyses parameters that are related to different physical properties of the 
galaxies: luminosity, star formation rate, light distribution inside the 
galaxies and the dominant stellar populations. The galaxy parameters we have 
focused our study on are: 
\begin{enumerate}
\item $^{0.1}r-$band absolute magnitude, $M_{^{0.1}r}$.
\item $^{0.1}(u-r)$ and $^{0.1}(g-r)$ colours.
\item The mono-parametric spectral classification based on the eigentemplates
expansion of galaxy's spectrum ${\rm eclass}$. This parameter ranges from about
$-0.35$ for early-type galaxies to $0.5$ for late-type galaxies 
(\citealt{yip04}). 
\item $r-$band surface brightness, $\mu_{50}$, computed inside the radius that
encloses 50\% of Petrosian flux, $r_{50}$.
\item  $r-$band concentration parameter defined as the ratio between the radii
that enclose 90\%  and 50\% of the Petrosian flux, $C=r_{90}/r_{50}$. 
Typically, early-type galaxies have $C>2.5$, while for late-types $C<2.5$ 
\citep{st01}. 
\item The S\'ersic index $n$ (taken from \citealt{blanton05}).
\end{enumerate}

It should be noted that in our samples we do not have galaxies with 
$r_{50}<2\arcsec$, i.e. $0.5\arcsec$ greater than the average seeing in SDSS 
(the average seeing in SDSS is below a conservative value of $1.5\arcsec$), 
therefore those galaxy parameters that involve any measure of the galaxy size 
should not be affected by the effect of seeing.

\begin{figure}
\includegraphics[width=90mm]{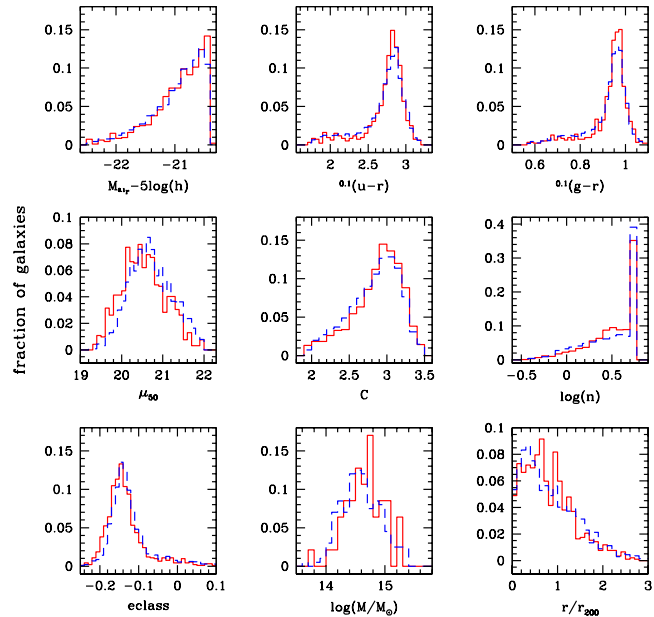}
\caption{The distributions of galaxy properties in our samples. Dashed blue 
line: C-K07-I sample, continuous red line: C-P04-I sample. We also show the 
distribution of cluster mass and the distribution of the projected 
cluster-centric distance in units of $r_{200}$.}
\label{fig1}
\end{figure}

We have introduced some further cut-offs in these galaxy properties besides 
luminosity, a complete list is shown in Table \ref{cutoffs}. This excludes only
a few galaxies from our analyses but is necessary in order to properly bin 
these quantities in the statistics we perform in next section. It should be 
mentioned that we do not introduce further cut-offs in the cluster mass nor in the 
cluster-centric distance. In Figure \ref{fig1} we show the distributions of 
galaxy parameters. 
As can be seen in Table \ref{tb:mean} and Figure \ref{fig1}, despite the 
clusters derived from the Popesso sample are, on average, slightly more massive
than those drawn from the Koester sample, galaxy properties in the C-P04-I 
sample are very similar to those in C-K07-I having the latter a slightly 
smaller  fraction of red galaxies (85\% and 89\% of galaxies redder than 
$g-r=0.82$, respectively).
\section{Which galaxy property correlates best with the environment?}
In order to determine which galaxy properties are more correlated with, either,
cluster mass or the projected distance to the centre of the cluster, we perform
here the same analysis carried out by \citet{blanton05} and \citet{hm2}.  
Details of how to compute the quantities $\sigma_X$ and $\sigma_{XY}$ are given
in \citet{hm2}, thus, we will just briefly summarise here their meaning. The 
galaxy property $X$ which correlates best with a quantity $\delta$ that 
characterises the environment (either, cluster mass or cluster-centric distance
in this work), is the one that minimises the variance of $\delta$ after 
subtracting the global trend of $\delta$ as a function of $X$. That is, the 
property $X$ that minimises the expression:
\begin{equation}
\sigma^2_X=\frac{1}{n-1}\sum_{j=1}^m 
\sum_{|X_i-\overline{X}_j|\leq\Delta X}(\delta_i-\overline{\delta}_j)^2, 
\label{sigmaX}
\end{equation}
in which $X$ has been binned into $m$ bins $2\Delta X$ wide and centred in 
$\overline{X}_j$ ($j=1,...,m$) and for each of these bins the mean value of 
$\delta$ is $\overline{\delta}_j$. Clearly, for any property $X$, the quantity
$\sigma_X$ will be smaller than the corresponding variance of $\delta$ with no
trend subtraction. We label this variance $\sigma$ and present all of our 
results as the difference $\sigma_X^2-\sigma^2$.

The quantity $\sigma_X$ is independent of the units of the physical quantity 
$X$, but it can be sensitive to the choice of binning. To have robust results 
we take care that each bin is larger than the mean errors in the considered 
parameter, is smaller than the features in the parameter's distribution, and 
contains a large enough number of galaxies. It can be straightforwardly 
generalised to two properties $X$ and $Y$ if one wants to analyse which pair 
of properties is most closely correlated with mass \citep{blanton05}. However 
we do not attempt to do so in this work since we do not have enough objects to
split them into as many bins as would be required for a proper computation of 
$\sigma_{XY}^2$ while still obtaining a reliable outcome.
\begin{table}
\center
\begin{tabular}{lcccc}
\hline
             & \multicolumn{2}{c}{C-K07-I sample} & \multicolumn{2}{c}{C-P04-I sample} \\
\hline
Property $X$ & $\sigma_X^2-\sigma^2$ & Significance& $\sigma_X^2-\sigma^2$ &Significance\\
\hline
$M_{r}^{0.1}$ & $-4$	    &50\%& $-4$        &45\%    \\
$(u-r)^{0.1}$ & $-3$ 	    &49\%& \fbox{$-8$} &49\%    \\
$(g-r)^{0.1}$ & $-1$        &46\%& $-3$        &44\%    \\
$\mu_{50}$    & $-3$        &48\%& $-6$        &47\%    \\
$C$           & \fbox{$-5$} &53\%& $-7$	     &49\%    \\
$\log(n)$     & $-1$        &45\%& $ 0$	     &41\%    \\
$eclass$      & $-3$ 	    &49\%& $-6$	     &48\%    \\
\hline
\end{tabular}
\label{tabM}
\caption{Galaxy parameters as cluster mass indicators, i.e., in this case 
$\delta\equiv\log_{10}(M)$ (see text for details). Quoted $\sigma^2$ values are
expressed in units of $10^{-4}$. Boxes highlight the lowest values of 
$\sigma_X^2-\sigma^2$, i.e., those corresponding to the galaxy parameter that 
predicts best the mass of the clusters. Quoted significances are assessed using
the bootstrap technique as described in the text.}
\end{table}
\subsection{Cluster mass}
\begin{figure}
\includegraphics[width=90mm]{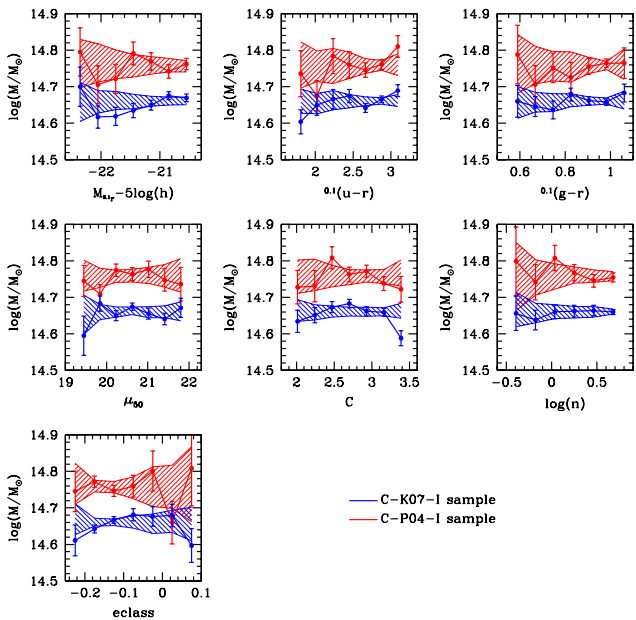}
\caption{The mean value of cluster mass as a function of the galaxy properties 
considered in our analysis. Error bars are errors in the mean value obtained by
the bootstrap re-sampling technique. Shaded areas correspond to the mean value 
obtained in the bootstrap re-samplings plus/minus 1$\sigma$ error bar.}
\label{figM}
\end{figure}
In this subsection we focus on how the galaxy properties relate to cluster 
mass. There is evidence that for groups of galaxies, galaxy colour is, among a
set of galaxy properties similar to the ones considered here, the property that correlates best with the mass of the system \citep{hm2,weinmann06}. It is 
interesting to test whether this is also true for more massive systems. We show
in figure \ref{figM} the mean mass of the clusters as a function of galaxy 
properties. There are some hints of what we would expect, for instance a redder mean colour of galaxies with increasing mass due to the higher fraction of red
galaxies. But the trends are not as well defined and strong as one may think 
{\em a priori}. Shaded areas in figure \ref{figM} correspond to the mean value
plus/minus 1 standard deviation error bar for 500 bootstrap re-samplings in 
which we assign to each galaxy the mass of the cluster to which another, 
randomly chosen, galaxy in the sample belongs to. It is clear that the trends 
are, with the exception of a few points, contained in the shaded areas. We 
should keep in mind that we are dealing here with bright galaxies, that in 
order to have volume limited samples we have only galaxies brighter than 
$M^{\ast}$. In Table \ref{tabM} we list the values of the differences 
$\sigma_X^2-\sigma^2$ for our samples of galaxies in clusters. Results differ 
from one sample to another. They do not even agree in the single property most
predictive of mass, let alone the second or the third.

The $\sigma_X$ analysis always provides a parameter that predicts best a given 
environment. Nevertheless, it does not mean that this parameter is good in 
predicting the environment. In order to test the significance of the results 
quoted in Table 3, we again use the bootstrap re-samplings to compute different
values of $\sigma_X$, that we label as $\sigma_X^b$, and obtain a distribution
for $\sigma_X^b$. We then compute the fraction, $F$, of re-samplings that gave
values $\sigma_X^b<\sigma_X$. The significance level of the measured $\sigma_X$
is then $1-F$. We list these significance values in Table \ref{tabM}. They 
confirm what is observed in figure \ref{figM}, that none of the galaxy 
properties correlates significantly with cluster mass. This may be interpreted 
as if bright galaxies in clusters were similar from cluster to cluster 
irrespective of cluster mass.
\begin{figure}
\includegraphics[width=90mm]{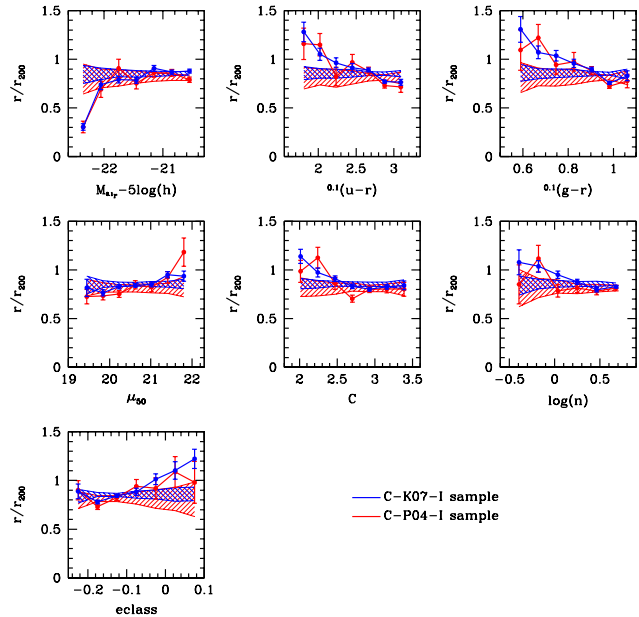}
\caption{The mean value of cluster-centric distance in units of $r_{200}$ as a
function of the galaxy properties considered in our analysis. Error bars are 
errors in the mean value obtained by the bootstrap re-sampling technique. 
Shaded areas correspond to the mean value obtained in the bootstrap 
re-samplings plus/minus 1$\sigma$ error bar.}
\label{figR}
\end{figure}
\subsection{Cluster-centric distance}
It is well known that spatial segregation occurs in galaxy clusters. We now 
compute $\sigma_X$ of the galaxies projected distance to the cluster centre 
(in units of $r_{200}$) as a function of the galaxy properties. For this 
purpose, we assume that the cluster centres are those determined by 
\citet{coenda08} as explained in section 2.1. In Figure \ref{figR} we show the
mean values of $r/r_{200}$ as a function of the galaxy parameters, as well as 
the corresponding areas defined by the bootstrap re-samplings. In contrast to 
what we found for the cluster mass in the figure \ref{figM}, there are clear 
trends of $r/r_{200}$ that show what is well known, earlier galaxies inhabit 
the inner regions of the clusters. The resulting values for $\sigma_X^2$ and 
their significance are quoted in table \ref{tabR}. 

Now we do find agreement between the two samples for the parameter that 
correlates best with cluster-centric distance. For both samples the $g-r$ 
colour ranks first. The colour $u-r$ is as good as $g-r$ in the C-P04-I sample,
while in the C-K07-I sample comes second, but by a small difference. From the 
third position onwards rankings differ. The significance is greater than 68\% 
only for 3 properties: the two colours and eclass. This result is consistent 
with \citet{hm2}, who found that the $g-r$ colour is the single parameter that
correlates best with the distance from the centre of the system in groups of 
galaxies.

In the upper left panel of figure \ref{figR} the point corresponding to the 
highest luminosities considered is well below the shaded area in both samples. 
We tested whether this is due to the brightest galaxy (BCG) alone by repeating 
the computations excluding the brightest galaxy of each cluster. In the 
brightest luminosity bin, 88\% and 60\% of the galaxies are BCGs for the 
C-K07-I and the C-P04-I samples, respectively. The resulting trend for the 
C-K07-I sample is kept almost unchanged, while for the C-P04-I sample the 
point lifts into the shaded area. However, it is important to take into account
that the actual brightest galaxy of each cluster is not always included in the 
samples (this is explored in detail in \citealt{coenda08}). 
\section{Discussion and conclusions}
In this paper we have extended the analyses by \citet{blanton05} and 
\citet{hm2} to galaxies in clusters. We analyse which galaxy properties 
correlate best with environment characterised by either cluster mass or 
cluster-centric distance. We use a sample of bright ($L>L^{\ast}$) galaxies in clusters 
of galaxies in SDSS identified by two different criteria: X-ray selected 
clusters and clusters selected according to their red sequence. The two 
sub-samples of clusters have some differences in the mean properties. X-ray 
selected clusters tend to have a slightly higher mean mass and a higher 
fraction of red galaxies than maxBCG systems.

We find that the properties of bright galaxies do not clearly depend on cluster
mass for 
systems more massive than $M\sim10^{14}M_{\odot}$. Although the mass range of 
our sample of cluster is not very large, the lack of dependence between mass 
and galaxy properties can be interpreted in terms of galaxy evolution. For 
systems with masses between $10^{14}M_{\odot}$ and $10^{15}M_{\odot}$ bright 
galaxies have experienced the same physical processes and therefore have 
similar properties. This result is consistent with \citet{hansen07}. Using 
clusters and groups identified in SDSS with the MaxBCG finder of K07, these 
authors find that above a cluster mass $\sim10^{14}M_{\odot}$ the fraction of 
red galaxies increases weakly with increasing mass. For bright galaxies, the 
lack of a significant correlation between some galaxy properties such as colour
and concentration with halo mass for halo masses above $\sim 10^{14}M_{\odot}$
is also present in \citet{weinmann06} (their figure 11). 
To check consistency with their findings we have computed 
the median of $^{0.1}(g-r)$ colour and the median of the 
concentration parameter as a function of mass for our samples. We have found 
that they are remarkably independent on cluster mass, taking 
values $\sim0.95$ and $\sim3.0$, respectively, in fully agreement with the 
higher mass bins of \citet{weinmann06}. 
For lower mass systems, the importance of processes like high 
speed encounters or those produced by 
the intra-cluster medium, tends to change faster with mass. 
\citet{yo02} and \citet{hm2} 
analysed groups of galaxies and found a clear dependence between mass and 
galaxy properties that tends to flatten as the mass of the systems grows. The 
higher masses considered by these authors are similar to the lower masses 
analysed in this work.
All these results seem to imply that above a mass $\sim10^{14}M_{\odot}$
clusters have a similar population of bright ($L>L^{\ast}$) galaxies, that
arises as a result of the action of the same physical mechanisms with similar
relative impact.

On the other hand, we find, as expected, that galaxy properties do correlate 
with cluster-centric distance. The property most affected by the 
cluster-centric distance is $g-r$ colour, followed closely by the $u-r$ colour.
These results are irrespective of the cluster selection criteria. This is also 
in agreement with \citet{hansen07}. Moreover, the parameter that best 
predicts the cluster-centric distance is the same found by \citet{hm2} for
groups of galaxies and \citet{blanton05} for the local density of field 
galaxies as the most predictive of environment. 

Galaxy parameters considered in this work can be classified into two classes. 
Those related to the physical properties of stars and those associated with the
light distribution. In the first set are colour, absolute magnitude and 
spectral type. To the second group belong the concentration parameter, the 
surface brightness and the S\'ersic index. Colour and spectral types of 
galaxies strongly depend on the age and metallicity of the stars as well as 
the present and recent star formation history. The luminosity of galaxies also 
depends on the same properties, although in a more indirect way. The fact that
from field to massive clusters colour is the most sensitive property of 
galaxies to the present time environment, suggests that what really matters is the 
overall evolution of the environment where a galaxy and its progenitors 
form the stars. Moreover, according to \citet{blanton05} and \citet{hm2}, the 
other two parameters that appear as one of the pair of properties most 
predictive of the environment for field and group's galaxies are the magnitude 
and the spectral type, both belonging to the first group of parameters. 

Among 
the phenomena that can affect the star formation history, we can mention the 
suppression or stimulation of the star formation due to interactions with the 
intergalactic medium or with other galaxies. Also, the past merger history of 
galaxies can play a fundamental role. In the particular case of the cluster 
environment, a natural segregation in the colour of galaxies arises as a 
result of cluster-centric gradient in the age of the stellar population as is 
observed in numerical simulations (e.g. \citealt{gao}). Galaxies in the inner 
regions of clusters would have older stellar populations and therefore would be
redder. These galaxies would also have had a longer time to deplete their gas 
reservoirs thus stopping star formation. In this scenario, the reddening as a
function of radius emerges naturally. 
However, these processes might suffer a saturation resulting in a flattening 
of the mean colour as a function of mass. 

The lack of a significant correlation between environment and the galaxy properties
in the second set described above, may imply that phenomena 
like ram pressure stripping or galaxy harassment would have had a secondary 
role in the evolution of bright galaxies.
Nevertheless, these physical processes would also have an impact in the first 
set of parameters. For example, ram pressure striping can 
remove an important fraction of the intragalactic gas producing a reddening 
of galaxies as a consequence of the reduction of the star formation rate. 
Numerical simulations show that galaxies can lose a high fraction of the gas 
after a single passage through the inner regions of a cluster (see for instance
\citealt{abadi99}). Therefore, above a certain mass ($\sim10^{14}M_{\odot}$), 
galaxies will experience the same physical processes acting with similar
relative effectiveness thus producing a saturation in the mass-colour 
relationship.
\begin{table}
\center
\begin{tabular}{lcccc}
\hline
             & \multicolumn{2}{c}{C-K07-I sample} & \multicolumn{2}{c}{C-P04-I sample} \\
\hline
Property $X$ & $\sigma_X^2-\sigma^2$ & Significance& $\sigma_X^2-\sigma^2$ &Significance\\
\hline
$M_{r}^{0.1}$ & $-7$	       &67\%&  $-7$         &50\%    \\
$(u-r)^{0.1}$ & $-13$ 	       &84\%&  \fbox{$-16$} &81\%    \\
$(g-r)^{0.1}$ & $\fbox{$-14$}$ &85\%&  \fbox{$-16$} &79\%    \\
$\mu_{50}$    & $-2$           &55\%&  $-8$         &61\%    \\
$C$           & $-5$           &63\%&  $-9$	  &64\%    \\
$\log(n)$     & $-4$           &60\%&  $-3$	  &49\%    \\
$eclass$      & $-8$ 	       &70\%&  $-7$	  &69\%    \\
\hline
\end{tabular}
\label{tabR}
\caption{Galaxy parameters as cluster-centric distance indicators, i.e., in 
this case $\delta\equiv r/r_{200}$ (see text for details). Quoted $\sigma^2$ 
values are expressed in units of $10^{-3}$. Boxes highlight the lowest values
of $\sigma_X^2-\sigma^2$, i.e., those corresponding to the galaxy parameter 
that predicts best the cluster-centric distance. Quoted significances are 
assessed using the bootstrap technique as described in the text.}
\end{table}
\section*{Acknowledgements}
We thank the anonymous referee for comments and suggestions that have
improved this paper.
HJM acknowledges the support of a Young Researcher's grant from Agencia 
Nacional de Promoci\'on Cient\'\i fica y Tecnol\'ogica Argentina, PICT 
2005/38087. This work has been partially supported with grants from Consejo 
Nacional de Investigaciones Cient\'\i ficas y T\'ecnicas de la Rep\'ublica 
Argentina (CONICET) and Secretar\'\i a de Ciencia y Tecnolog\'\i a de la 
Universidad de C\'ordoba.

Funding for the Sloan Digital Sky Survey (SDSS) has been provided by the 
Alfred P. Sloan Foundation, the Participating Institutions, the National 
Aeronautics and Space Administration, the National Science Foundation, the U.S.
Department of Energy, the Japanese Monbukagakusho, and the Max Planck Society. The SDSS Web site is http://www.sdss.org/. The SDSS is managed by the 
Astrophysical Research Consortium (ARC) for the Participating Institutions. 
The Participating Institutions are The University of Chicago, Fermilab, the 
Institute for Advanced Study, the Japan Participation Group, The Johns Hopkins
University, the Korean Scientist Group, Los Alamos National Laboratory, the 
Max Planck Institut f\"ur Astronomie (MPIA), the Max Planck Institut f\"ur 
Astrophysik (MPA), New Mexico State University, University of Pittsburgh, 
University of Portsmouth, Princeton University, the United States Naval 
Observatory, and the University of Washington.

\label{lastpage}
\end{document}